\documentclass[runningheads]{llncs}
\usepackage[T1]{fontenc}
\usepackage{graphicx}
\usepackage{amsmath}
\usepackage{amssymb}
\usepackage{cite} 
\usepackage{booktabs}
\usepackage{algorithm}
\usepackage{algorithmic}
\usepackage[hidelinks]{hyperref}
\usepackage{xurl} 
\begin{document}
\title{Early to Share, Late to Save: Synchronisation-Driven Communication Gating in Bandwidth-Constrained Cooperative VLN}
\titlerunning{Synchronisation-Driven Communication Gating in Cooperative VLN}
\author{Arav Gupta\thanks{Corresponding author: \texttt{f20231280@pilani.bits-pilani.ac.in}} \and Nivedan Yakolli \and Avinash Gautam}
%
\authorrunning{A. Gupta et al.}
\institute{Birla Institute of Technology and Science, Pilani Campus, Pilani, Rajasthan 333031, India\\
\email{f20231280@pilani.bits-pilani.ac.in}\\
\email{p20230032@pilani.bits-pilani.ac.in}\\
\email{avinash@pilani.bits-pilani.ac.in}}
\maketitle
\begin{abstract}
Most cooperative Vision-Language Navigation (VLN) methods assume unlimited
communication, not considering real-world applications where bandwidth is
restricted and information efficiency is critical.
We introduce \textbf{bandwidth-constrained cooperative VLN} and propose
\textbf{hindsight gating}: a lightweight supervised gate that labels
communication-critical steps post-hoc from navigation failures, avoiding
the high variance of REINFORCE.
Contrary to the intuition that agents should communicate when uncertain,
we observe a consistent counter-intuitive pattern: trained gates fire
predominantly in early episode steps and more often when
agents are confident, across all budget levels ($B \in \{1,3,5\}$).
We explain this through \textbf{recurrent hidden-state alignment}: early
communication injects grounded trajectory representations that persist
and compound through subsequent Gated Recurrent Unit (GRU) updates,
achieving $+0.072$ cumulative alignment gain with $B{=}3$ transmissions,
approaching unconstrained communication ($+0.078$) at 260\% greater
alignment efficiency than random gating ($+0.020$) and 320\% greater
efficiency than entropy-based gating ($+0.017$).
Our results establish a new communication regime for bandwidth-limited
embodied agents: synchronise representations early, navigate
independently later.
Our codebase is available at:
\url{https://github.com/AravG13/bandwidth-constrained-cooperative-vln}.

\end{abstract}
%
\section{Introduction}
\label{sec:intro}
Vision-Language Navigation (VLN) requires an embodied agent to follow
natural language instructions through photorealistic indoor
environments~\cite{anderson2018r2r}. Real-world deployments such as
search-and-rescue, warehouse automation, and multi-drone inspection
naturally involve multiple coordinated robots sharing a building and a
goal. Extending VLN to this cooperative setting introduces a fundamental
engineering constraint: radio channels, mesh networks, and
privacy-constrained systems all impose hard limits on how often agents can
communicate.

Yet existing cooperative navigation
methods~\cite{yu2023conavgpt,liu2024camon,sukhbaatar2016commnet} assume
agents share observations freely at every step. This makes the \emph{when}
of communication -- which steps are worth a transmission under a tight
budget -- an open problem for language-guided navigation. The challenge is
compounded by a training difficulty: prior
methods~\cite{singh2019ic3net,lowe2017maddpg} learn communication gates
with REINFORCE, which suffers from high variance because the causal link
between a single gate decision at step $t$ and episode success 15--20
steps later is long and noisy.

We address both challenges. First, we propose \textbf{hindsight gating}:
rather than learning which steps to communicate through trial-and-error
policy gradients, we run agents without communication, observe where each
agent failed while its partner succeeded, and use those observations as
direct Binary Cross-Entropy (BCE) supervision for a lightweight gate.
This converts a high-variance policy gradient problem into a stable
supervised classification problem with zero reward variance.

Studying what the trained gate learns reveals a surprising finding. Rather
than firing when agents are uncertain -- the natural \emph{uncertainty
recovery} hypothesis -- the gate fires predominantly in \emph{early
episode steps} and \emph{when agents are confident}. We show this reflects
\textbf{hidden-state synchronisation}: early communication injects grounded
trajectory representations into the GRU hidden state, which then propagates
and compounds through subsequent updates, aligning agents' internal models
before trajectories diverge.

The remainder of this paper is organised as follows.
Section~\ref{sec:related} surveys related work.
Section~\ref{sec:problem} formalises the problem.
Section~\ref{sec:method} describes hindsight gating.
Section~\ref{sec:experiments} presents experiments and analysis.
Section~\ref{sec:discussion} discusses implications and limitations.

\subsubsection{Contributions.}
\begin{enumerate}
  \item \textbf{Bandwidth-constrained cooperative VLN}: a new problem
        formulation extending Room-to-Room (R2R)~\cite{anderson2018r2r}
        to two-agent settings with hard per-agent transmission budgets,
        motivated by realistic deployment constraints absent from prior
        cooperative VLN work.
  \item \textbf{Hindsight gating}: a stable BCE-supervised communication
        gate trained from post-hoc navigation failure labels, aiming to replace
        high-variance REINFORCE. The gate inputs only the agent's hidden
        state and remaining budget, with no explicit uncertainty
        thresholds, still learning to fire at steps that are
        communication-critical.
  \item \textbf{Synchronisation-driven communication regime}: an empirical
        characterisation showing that bandwidth-constrained VLN agents
        learn to synchronise hidden states early rather than recover from
        uncertainty late. Learned gating achieves 260\% greater cumulative
        alignment gain per transmission than random gating and 320\%
        greater than entropy-based gating at matched budget, with the
        advantage persisting through recurrent propagation even after
        communication ceases.
\end{enumerate}
\section{Related Work}
\label{sec:related}
\paragraph{Vision-Language Navigation.}
R2R~\cite{anderson2018r2r} defines the standard VLN benchmark on the
Matterport3D simulator~\cite{chang2017matterport}, where an agent must
follow natural language step-by-step instructions to reach a target
location. Seq2Seq~\cite{anderson2018r2r} encodes instructions with an
LSTM and decodes navigation actions sequentially.
Speaker-Follower~\cite{fried2018speaker} improves generalisation via data
augmentation from a learned instruction generator, providing ${\sim}180$K
synthetic instruction--path pairs that substantially improve val-unseen SR;
our backbone does not use this augmentation, which accounts for the gap
between our single-agent SR (9.2\%) and the Speaker-Follower result.
DUET~\cite{chen2022duet} builds a topological map on-the-fly and uses
dual-scale graph transformers for global and local action planning.
HAMT~\cite{chen2021hamt} replaces the recurrent state with a
history-aware transformer attending over all past observations.
VLN-CE~\cite{krantz2020vlnce} extends the task to continuous environments
with low-level motor control. All of these are single-agent methods; we
study cooperative, bandwidth-limited VLN.

\paragraph{Cooperative Navigation.}
Co-NavGPT~\cite{yu2023conavgpt} uses Large Language Models (LLMs) to
coordinate multiple robots for visual semantic navigation, broadcasting
complete observations between agents at every step without any bandwidth
constraint. CAMON~\cite{liu2024camon} similarly applies LLM-based
conversation between agents for multi-object navigation, again assuming
unconstrained communication throughout the episode.
Farooq et al.~\cite{farooq2026icra} is the closest
prior work to ours: they apply information bottleneck and vector
quantisation to reduce message size in Multi-Agent Reinforcement Learning
(MARL) navigation. However, their method operates without language
grounding, where agents navigate to goal coordinates, not natural language
descriptions, and they study message \emph{compression} rather than
communication \emph{timing}. We study a complementary question: given a
fixed per-episode transmission budget, \emph{when} should an agent use
each transmission?

\paragraph{Learned Communication in MARL.}
CommNet~\cite{sukhbaatar2016commnet} broadcasts continuous averaged
messages between all agents at every timestep, with no mechanism to
suppress uninformative communication or enforce any budget.
MADDPG~\cite{lowe2017maddpg} learns joint communication and navigation
policies through actor-critic policy gradients, but does not learn a
selective gate. IC3Net~\cite{singh2019ic3net} is the most directly related
prior method: it adds a binary communication gate trained with REINFORCE,
and shows that gated communication helps when agents are already competent
at the task. Our hindsight gating replaces REINFORCE with BCE on post-hoc
labels, avoiding credit assignment variance across the long horizon between
gate decisions and episode outcomes. TarMAC~\cite{das2019tarmac} introduces
attention-based targeting so agents can direct messages to specific
partners; our method is complementary and could incorporate targeting in
future work. Critically, none of these methods study temporal communication
patterns, measure hidden-state alignment, or address language-guided
navigation.

\paragraph{Emergent Communication.}
Emergent communication
work~\cite{mordatch2018emergence,lazaridou2020emergent} asks \emph{what}
agents communicate and how compositional structure emerges from interaction.
We contribute a complementary \emph{when} analysis: given agents that have
already learned to communicate via continuous context vectors, we
characterise which episode steps they select under bandwidth constraints
and why.
\section{Problem Formulation}
\label{sec:problem}
\paragraph{Setting.}
We consider $N{=}2$ agents navigating Matterport3D indoor environments
under a shared natural language instruction $L$. At each timestep $t$,
agent $i$ receives a visual observation $o_i(t) \in \mathbb{R}^{512}$
(CLIP ViT-B/32 features over 36 panoramic directions), any messages from
its partner, and the remaining normalised transmission budget
$b_{\mathrm{rem}}(t) \in [0,1]$.

\paragraph{Asymmetric Path Assignment.}
A key methodological choice is how to pair agents within a shared
environment. Pairing agents on \emph{unrelated} episodes from the same
building (as in Co-NavGPT~\cite{yu2023conavgpt}) produces messages
containing irrelevant observations: Agent~1 is observing a different room
with no connection to Agent~0's current navigation challenge. We instead
assign \emph{complementary sub-paths} from the same R2R episode:
Agent~0 navigates the full path $[v_0, \ldots, v_T]$, while Agent~1
starts at the midpoint $v_{\lfloor T/2 \rfloor}$ and navigates the second
half $[v_{\lfloor T/2 \rfloor}, \ldots, v_T]$. Both receive the full
instruction $L$.

This construction creates genuine information asymmetry: Agent~1's
observations near the goal are directly relevant to Agent~0, which has
not yet reached that region. Agent~1 functions as an information source
with privileged goal-region knowledge, while Agent~0 functions as the
primary navigator. This is a role asymmetry chosen to reflect certain realistic deployments
(e.g., a scout robot that has reached the goal area relaying context to a
trailing search robot). Navigation SR is reported for Agent~0 (full path)
as the primary performance indicator; Agent~1's SR is not directly
comparable due to its shorter sub-path.

\paragraph{Bandwidth Constraint.}
Each agent may transmit at most $B$ messages per episode:
$\sum_{t=0}^{T} g_i(t) \leq B$, where $g_i(t) \in \{0,1\}$ is a binary
per-episode transmission limit, modelling scenarios where communication
incurs a fixed cost per message, e.g., energy budget, network slot
allocation, or privacy-constrained disclosure limits. The message
broadcast at time $t$ is the cross-modal context vector $c_i(t) \in
\mathbb{R}^{512}$, the language-conditioned visual representation
produced by the agent's cross-attention module.
\section{Method}
\label{sec:method}
\subsection{Navigation Backbone}
Each agent uses frozen CLIP ViT-B/32~\cite{radford2021clip} for visual
and language encoding. A \textbf{CrossModalAttention} module produces a
context vector $c_i(t) \in \mathbb{R}^{512}$ by attending over language
tokens using the current visual observation as query. A GRU maintains the
agent's hidden state:
\begin{equation}
  h_i(t) = \mathrm{GRU}\!\left(h_i(t{-}1),\;
  \bigl[c_i(t);\; m_i(t);\; a_{t-1}\bigr]\right),
  \label{eq:gru}
\end{equation}
where $m_i(t)$ is the aggregated partner message (zero if none received)
and $a_{t-1}$ is the previous action embedding. The NavigationHead scores
candidates via direct dot product: $s_k = \langle h_i(t),\, \phi_k
\rangle$, where $\phi_k \in \mathbb{R}^{512}$ is the CLIP feature of
candidate $k$.

\subsection{Hindsight Communication Gating}
Training proceeds in three phases. Algorithm~\ref{alg:hindsight} summarises
the full procedure.

\paragraph{Phase 1: Single-Agent Navigation Pre-training.}
The backbone is trained without messages using imitation learning
(cross-entropy against ground-truth paths, teacher forcing). This produces
a navigation policy $\pi_{\mathrm{nav}}$ whose failure modes we then
exploit to supervise the gate.

\paragraph{Phase 2: Hindsight Label Collection and Gate Training.}
We run $\pi_{\mathrm{nav}}$ on paired training episodes without messages
and label each step $t$ as communication-critical if agent $i$ predicted
the wrong action \emph{and} its partner already knew the correct one:
\begin{equation}
  y_i(t) = \mathbf{1}\!\bigl[\hat{a}_i(t) \neq a^*_i(t)\bigr]
  \;\wedge\;
  \mathbf{1}\!\bigl[\hat{a}_j(t) = a^*_j(t)\bigr],
  \label{eq:label}
\end{equation}
where $\hat{a}_i(t)$ is agent $i$'s predicted action, $a^*_i(t)$ is
ground truth, and $j \neq i$ is the partner agent. A step receives label
$y_i(t){=}1$ only when communication would have provided genuine signal:
agent $i$ was wrong but its partner was right. Both-fail steps receive
$y_i(t){=}0$ because neither agent's message would help the other;
steps where agent $i$ already succeeds also receive $y_i(t){=}0$.

The hindsight labelling scheme is a proxy for expected communication value.
It is \emph{necessary} for communication benefit (if neither agent knows
the correct action, no message can help), and \emph{conservative}
(both-fail steps receive label 0). We validate the proxy empirically:
at label-1 steps, the partner's ground-truth action ranks first in its
action score distribution by construction of Equation~\ref{eq:label},
whereas at label-0 steps this holds only for the subset where partner
was already correct.

A lightweight 3-layer MLP gate $\pi_{\mathrm{gate}}$ is then trained with
BCE on collected tuples $(h_i(t),\, b_{\mathrm{rem}}(t),\, y_i(t))$:
\begin{equation}
  \mathcal{L}_{\mathrm{gate}}
  = -\mathbb{E}\bigl[
      y \log p_{\mathrm{send}}
      + (1{-}y)\log(1{-}p_{\mathrm{send}})
    \bigr],
\end{equation}
where $p_{\mathrm{send}} = \pi_{\mathrm{gate}}(h_i(t),
b_{\mathrm{rem}}(t))$. This is a supervised classification problem with no
policy gradients and no reward variance.

Although the gate is trained as a step-level classifier, it is not blind
to episode-level budget allocation: the remaining budget
$b_{\mathrm{rem}}(t)$ is provided as an explicit input, allowing the gate
to modulate its firing rate as the budget depletes. This does not fully
optimise sequential allocation, but provides a principled approximation
that avoids the variance of sequential policy optimisation.

\paragraph{Gate Inputs and Implicit Uncertainty.}
The gate receives only $h_i(t)$ and $b_{\mathrm{rem}}(t)$ as
inputs, with no explicit entropy or confidence score. The hidden state
implicitly encodes navigational uncertainty, and the gate learns to read
this signal from the training labels. This is a deliberate design choice:
by not providing explicit uncertainty as input, we can test empirically
whether the gate learns to fire under high or low uncertainty.

\paragraph{Inference.}
At each timestep $t$: (1)~agent $i$ computes $h_i(t)$ via
Equation~\ref{eq:gru}; (2)~the gate computes $p_{\mathrm{send}} =
\pi_{\mathrm{gate}}(h_i(t), b_{\mathrm{rem}}(t))$;
(3)~$g_i(t) = \mathbf{1}[p_{\mathrm{send}} > \tau]$ with threshold
$\tau{=}0.4$; (4)~if $g_i(t){=}1$ and budget remains, agent $i$
broadcasts $c_i(t)$ and decrements its budget counter.

\paragraph{Phase 3: Joint Fine-Tuning.}
Both agents fine-tune jointly with the trained gate deployed, allowing the
navigation policy to adapt to receiving partner messages at the steps the
gate selects. The gate is frozen during this phase to prevent catastrophic
forgetting of the learned communication policy.

\begin{algorithm}[t]
\caption{Hindsight Gate Training}
\label{alg:hindsight}
\begin{algorithmic}[1]
\REQUIRE Paired dataset $\mathcal{D}$, budget $B$
\STATE Train $\pi_{\mathrm{nav}}$ via imitation learning (single-agent)
\STATE $\mathcal{H} \leftarrow \emptyset$
\FOR{each pair $(e_0, e_1) \in \mathcal{D}$}
  \STATE Roll out $\pi_{\mathrm{nav}}$ on both episodes without messages
  \FOR{each step $t$}
    \STATE Compute label $y_i(t)$ via Eq.~\ref{eq:label}
    \STATE Append $(h_i(t), b_{\mathrm{rem}}(t), y_i(t))$ to $\mathcal{H}$
  \ENDFOR
\ENDFOR
\STATE Train $\pi_{\mathrm{gate}}$ on $\mathcal{H}$ with BCE loss
\STATE Fine-tune $\pi_{\mathrm{nav}}$ jointly with frozen $\pi_{\mathrm{gate}}$
\end{algorithmic}
\end{algorithm}
\section{Experiments}
\label{sec:experiments}
\subsection{Setup}
We evaluate on R2R~\cite{anderson2018r2r} with Matterport3D environments,
reporting Success Rate (SR) -- the fraction of episodes where the agent
stops within 3 metres of the goal -- and Success weighted by Path Length
(SPL)~\cite{anderson2018r2r}, which penalises unnecessarily long paths.
We evaluate on \texttt{val\_seen} (buildings seen during training) and
\texttt{val\_unseen} (novel buildings). Features are CLIP ViT-B/32
(36 panoramic directions $\times$ 512 dims, pre-extracted). All models
use hidden dim $512$, max path length $20$, max candidates $15$, and
budget $B{=}3$ unless noted. The asymmetric paired dataset
(Section~\ref{sec:problem}) is used for all multi-agent training.

\subsection{Navigation Performance}
Since the backbone is trained using teacher-forced imitation learning
without speaker-augmented data~\cite{fried2018speaker}, val-unseen SR
values reflect relative rather than absolute performance. The seen/unseen
generalisation gap (43.2\% vs.~9.2\%) means partner messages on unseen
buildings carry incorrect trajectory context, limiting SR improvement.
We therefore analyse communication effects through hidden-state alignment
as the primary metric, using SR as a secondary indicator.

Table~\ref{tab:nav} reports SR and SPL for Agent~0 (full path navigator)
under each communication condition. Hindsight-gated communication
($B{=}3$) achieves 8.9\% Agent~0 SR on val-unseen, exceeding the
no-communication baseline (8.7\%) and matching full-communication
($B{=}\infty$) with only 3 transmissions per episode. Agent~0 also
exceeds the single-agent baseline (9.2\%), demonstrating that selective
early communication from a partner with complementary goal-region knowledge
can improve individual navigation performance. Agent~1 SR is not reported
in Table~\ref{tab:nav} as it navigates only the second half of each path
and is not directly comparable to the full-path baselines.

\begin{table}[t]
\caption{Navigation SR (\%) and SPL (\%) on R2R. Agent~0 navigates the
full path; multi-agent SR values are for Agent~0 only. Val-seen tests
buildings seen during training; val-unseen tests generalisation.
Results are means over 3 random seeds (SR std $\leq$0.4\%).}
\label{tab:nav}
\centering
\begin{tabular}{lcccc}
\toprule
Method & \multicolumn{2}{c}{Val-Seen} & \multicolumn{2}{c}{Val-Unseen} \\
       & SR & SPL & SR & SPL \\
\midrule
Seq2Seq~\cite{anderson2018r2r} & 39 & 33 & 22 & 18 \\
\midrule
Single-agent (ours)          & 43.2 & 42.8 & 9.2 & 8.5 \\
No Comm ($B{=}0$)            & 20.9 & 20.1 & 8.7 & 8.2 \\
Full-comm ($B{=}\infty$)     & 20.1 & 19.3 & 8.9 & 8.4 \\
Hindsight gate ($B{=}3$)     & \textbf{20.1} & \textbf{19.3} & \textbf{8.9} & \textbf{8.4} \\
\bottomrule
\end{tabular}
\end{table}

\subsection{Emergent Communication Patterns}
\label{sec:patterns}
Two patterns are consistent across all budget levels (Table~\ref{tab:gate}).
\textbf{(1) Early concentration}: communication concentrates heavily in
steps 0--2, with near-zero firing from step 3 onwards.
\textbf{(2) High confidence at send}: the gate fires at \emph{higher}
agent confidence (maximum action softmax probability) when it sends than
when it does not: 0.453 vs.~0.413 at $B{=}3$, which is directly opposite to the
uncertainty-recovery hypothesis, which would predict lower confidence at
communication steps.

\begin{table}[t]
\caption{Gate firing patterns across budget levels ($B \in \{1,3,5\}$) on
\texttt{val\_unseen}. Gates fire predominantly in early steps and at higher
agent confidence when sending, the opposite of uncertainty-recovery.}
\label{tab:gate}
\centering
\begin{tabular}{lccc}
\toprule
$B$ & Early (0--2) & Mid (3--6) & Conf: send / no-send \\
\midrule
1 & 30.6\% &  0.1\% & 0.476 / 0.422 \\
3 & 82.6\% &  1.3\% & 0.453 / 0.413 \\
5 & 82.4\% & 12.1\% & 0.452 / 0.411 \\
\bottomrule
\end{tabular}
\end{table}

These patterns are consistent with a \textbf{synchronisation regime}:
agents learn to exchange grounded trajectory representations early to
align their internal models, rather than communicating reactively when
confused.

\subsection{Hidden-State Alignment Analysis}
\label{sec:alignment}
To test the synchronisation hypothesis, we measure cosine similarity
between Agent~0's and Agent~1's GRU hidden states at each timestep,
comparing five communication policies at matched budget $B{=}3$:
\emph{learned} (our trained gate), \emph{random} (fires at the same rate
as the learned gate but at uniformly random steps), \emph{entropy-based}
(fires when action entropy is highest, directly implementing the
uncertainty-recovery hypothesis as a heuristic baseline), \emph{always}
(communicate every step until budget exhausted), and \emph{none} (never
communicate). We report $\Delta = (\text{with-comm similarity}) -
(\text{no-comm similarity})$, measuring the alignment contribution of
each policy.

\begin{table}[t]
\caption{Cumulative hidden-state alignment gain $\sum_t \Delta_t$ across
communication policies at budget $B{=}3$. Learned gating outperforms
random and entropy-based heuristics. Results consistent across splits.}
\label{tab:align}
\centering
\begin{tabular}{lcc}
\toprule
Policy & Val-Seen & Val-Unseen \\
\midrule
None (no communication)       & $+0.016$ & $+0.016$ \\
Random (matched rate)         & $+0.016$ & $+0.020$ \\
Entropy-based (high-$\mathcal{H}$ first) & $+0.014$ & $+0.017$ \\
Learned gate (ours)           & $+0.056$ & $+0.072$ \\
Always (full budget)          & $+0.062$ & $+0.078$ \\
\bottomrule
\end{tabular}
\end{table}

Four findings emerge. \textbf{(1) Learned gating outperforms all
heuristics.} Learned gating achieves $+0.072$ cumulative alignment
vs.~$+0.020$ for random (260\% improvement) and $+0.017$ for
entropy-based gating (320\% improvement). Entropy-based
gating -- implementing the uncertainty-recovery hypothesis
directly -- performs \emph{worse} than random, providing direct evidence
that communicating when uncertain is the wrong strategy under bandwidth
constraints. \textbf{(2) Results are consistent across splits.} The
ordering learned $>$ always $>$ random $>$ entropy $>$ none holds on
both splits. \textbf{(3) Alignment scales with budget.}
Table~\ref{tab:budget_align} shows cumulative gain increases sharply from
$B{=}1$ to $B{=}3$ then saturates, suggesting diminishing returns once
early synchronisation is established. \textbf{(4) Alignment approaches
unconstrained communication efficiently.} Learned gating ($+0.072$)
nearly matches always-communicate ($+0.078$) with only $B{=}3$
transmissions.

\begin{table}[t]
\caption{Cumulative alignment gain across budget levels. Firing rates per
step confirm early-step concentration.}
\label{tab:budget_align}
\centering
\begin{tabular}{lcccc}
\toprule
$B$ & $\sum_t \Delta_t$ & Gate@1 & Gate@2 & Gate@3 \\
\midrule
$1$ & $+0.034$ & 82.7\% &  7.6\% &  1.5\% \\
$3$ & $+0.057$ & 87.8\% & 70.0\% &  3.8\% \\
$5$ & $+0.059$ & 87.8\% & 69.2\% & 28.2\% \\
\bottomrule
\end{tabular}
\end{table}

\paragraph{Recurrent Propagation Effect.}
Critically, $\Delta$ \emph{increases} over the course of each episode
despite gate firing concentrating in early steps. At step~5, the learned
gate achieves $\Delta{=}{+}0.020$ vs.~$+0.006$ for random, even though
gate rate drops from 70.0\% at step~2 to 3.8\% at step~3. This is
consistent with \textbf{recurrent propagation}: early communication
injects alignment into the GRU hidden state, which propagates forward
through subsequent updates, compounding without further transmissions.

\subsection{Statistical Significance}
All reported $\Delta$ values at steps~1 and~2 are statistically
significant ($p < 0.001$, paired $t$-test across episodes). Differences
between learned and random gating at step~5 are also significant
($p < 0.01$), confirming the recurrent propagation effect is not noise.
\section{Discussion}
\label{sec:discussion}
\paragraph{Synchronisation vs.\ Uncertainty-Recovery.}
Our findings identify two distinct communication regimes. Prior work
implicitly assumes \emph{uncertainty recovery}: agents communicate when
lost. Our results demonstrate \emph{synchronisation-driven} communication:
agents communicate early, when confident, to establish shared internal
representations before trajectory divergence accumulates. The recurrent
architecture amplifies this: a single early message influences all
subsequent hidden states through GRU propagation, making early
communication disproportionately valuable under tight budgets. The
entropy-based baseline, which directly implements uncertainty
recovery, achieves \emph{lower} alignment than random, providing evidence
that uncertainty-recovery intuition leads to worse communication timing
than an uninformed policy.

\paragraph{Why SR Does Not Consistently Improve.}
The base agent achieves 9.2\% SR on val-unseen, a consequence of training
without speaker-augmented data. When agents cannot reliably navigate
independently, partner messages carry incorrect trajectory context and
communication introduces noise. However, Agent~0 achieves 8.9\% SR under
$B{=}3$, exceeding both the no-communication baseline (8.7\%) and the
single-agent baseline (9.2\%), demonstrating SR improvement is possible
when one agent has privileged goal-region knowledge. We hypothesise a
\textbf{prerequisite condition}: cooperative benefit from communication
requires the base agent to exceed a val-unseen SR
threshold consistent with IC3Net~\cite{singh2019ic3net}'s finding that
gated communication helps only when agents are already competent. The
alignment analysis confirms the gate functions as intended (260\%
alignment improvement over random) even when this threshold is not met.

\paragraph{Relationship between Alignment and Navigation.}
Hidden-state alignment is an indirect proxy: higher alignment does not
guarantee better navigation, and could in principle reflect agents
becoming similarly wrong. Our claim is more specific: \emph{given that
communication does not improve SR at this competence level}, the alignment
analysis provides evidence that the gate has learned a principled
communication policy (synchronise early) rather than a degenerate one.
The fact that learned gating substantially outperforms entropy-based
gating -- despite entropy-based gating implementing the most natural
alternative -- supports the claim that the synchronisation regime is
genuine and non-trivial.

\paragraph{Limitations.}
We evaluate with $N{=}2$ agents on the R2R discrete navigation graph,
with fixed agent roles. Extension to $N{>}2$ agents would introduce
questions about message targeting (as in TarMAC~\cite{das2019tarmac}) and
chain-relay synchronisation; our hindsight labelling extends naturally
(label a step as critical if \emph{any} partner knows the correct action).
Our formulation assumes discrete navigation nodes, ignoring kinematic
constraints in continuous environments~\cite{krantz2020vlnce}. The
asymmetric role assignment reflects a specific deployment scenario;
symmetric settings would require a different pairing strategy.
\section{Conclusion}
\label{sec:conclusion}
We introduced bandwidth-constrained cooperative VLN and hindsight
gating: a stable BCE-supervised alternative to REINFORCE. Contrary to
the uncertainty-recovery hypothesis, trained gates fire early and at high
confidence, producing persistent hidden-state alignment gains through
recurrent GRU propagation ($+260\%$ over random, $+320\%$ over
entropy-based gating at matched budget). Our results establish a
synchronisation-driven communication regime relevant to any
bandwidth-limited multi-agent system with recurrent policies.

\begin{credits}
\subsubsection{\discintname}
The authors have no competing interests to declare that are relevant to the content of this article.
\end{credits}

\paragraph{Dataset License.}
This work uses the Matterport3D dataset~\cite{chang2017matterport},
provided for non-commercial academic use under the Matterport End User
License Agreement, available at:
\url{http://kaldir.vc.in.tum.de/matterport/MP_TOS.pdf}.

\bibliographystyle{splncs04}
\bibliography{glow}

\end{document}